\documentclass{PoS}
\pdfoutput=1
\usepackage{amsmath,bbm}

\newcommand{\1}{\mathbbm{1}}
\newcommand{\eps}{\varepsilon}
\newcommand{\ev}[1]{\langle\smash{#1}\rangle}
\newcommand\SO{\text{SO}}
\newcommand\Sp{\text{Sp}}
\newcommand\SU{\text{SU}}
\newcommand\U{\text{U}}
\DeclareMathOperator{\diag}{diag}
\DeclareMathOperator{\re}{Re}
\DeclareMathOperator{\tr}{tr}

\title{Banks-Casher-type relations for complex Dirac
  spectra\thanks{Supported by RIKEN iTHES Project and JSPS KAKENHI
    Grants Number 25887014 (TK), by DFG (TW), and by Maryland Center
    for Fundamental Physics (NY).}}

\ShortTitle{Banks-Casher-type relations for complex Dirac spectra}

\author{Takuya Kanazawa\\
  Quantum Hadron Physics Laboratory, RIKEN, Saitama 351-0198, Japan\\
  E-mail: \email{takuya.kanazawa@riken.jp}}

\author{\speaker{Tilo Wettig}\\
  Department of Physics, University of Regensburg, 93040 Regensburg,
  Germany\\ 
  E-mail: \email{tilo.wettig@ur.de}}

\author{Naoki Yamamoto\\
  Department of Physics, University of Maryland, College Park, MD
  20742-4111, USA\\
  E-mail: \email{nyama@umd.edu}}

\abstract{For theories with a sign problem there is no analog of the
  Banks-Casher relation. This is true in particular for QCD at nonzero
  quark chemical potential. However, for QCD-like theories without a
  sign problem the Banks-Casher relation can be extended to the case
  of complex Dirac eigenvalues. We derive such extensions for the
  zero-temperature, high-density limits of two-color QCD, QCD at
  nonzero isospin chemical potential, and adjoint QCD. In all three
  cases the density of the complex Dirac eigenvalues at the origin is
  proportional to the BCS gap squared.}

\FullConference{31st International Symposium on Lattice Field Theory
  LATTICE 2013\\
   July 29 - August 3, 2013\\
    Mainz, Germany}

\begin{document}

\section{Introduction}

The Banks-Casher relation $|\ev{\bar\psi\psi}|=\pi\rho(0)$
\cite{Banks:1979yr}, where $\ev{\bar\psi\psi}$ is the chiral
condensate and $\rho(\lambda)$ is the Dirac eigenvalue density, links
the order parameter for spontaneous chiral symmetry breaking to an
accumulation of Dirac eigenvalues near zero.  Specifically, chiral
symmetry breaking occurs if the low-lying Dirac eigenvalues are of
order $1/V_4$, where $V_4$ is the space-time volume.  For the
derivation of the Banks-Casher relation it is essential that the
fermionic measure is positive definite \cite{Leutwyler:1992yt}.  If it
is not, as is the case in QCD at nonzero density, $\rho(0)$ is
undefined and the connection between $\ev{\bar\psi\psi}$ and the
(complex) Dirac spectrum is more complicated
\cite{Osborn:2005ss,Osborn:2008jp}.  Here, we consider QCD-like
theories at high density (and zero temperature) that have complex
Dirac spectra but positive definite fermionic measures.  We show that
Banks-Casher-type relations can be obtained for these theories, which
in all cases have the form
\begin{align}
  \label{eq:BC}
  \Delta^2=\frac{2\pi^3}{3d_\text{rep}}\rho(0)\,,
\end{align}
where $\Delta$ is the BCS gap, $\rho(\lambda)$ is now a
two-dimensional spectral density, and $d_\text{rep}$ is the dimension
of the color group representation in which the fermions transform.  To
derive these relations, we had to construct the corresponding
low-energy effective theories.  All details can be found in
\cite{Kanazawa:2012zr}.

\section{Derivation of Banks-Casher-type relations}

\subsection{Preliminaries}

The main idea of the derivation is to write down the partition
function $Z(M)$ as a function of the quark mass matrix $M$, both in
the fundamental (QCD-like) theory and in the corresponding effective
theory.  Taking suitable derivatives then yields $\sim\rho(0)$ in the
fundamental theory and $\sim\Delta^2$ in the effective theory.  The
Banks-Casher-type relation is obtained by identifying these results.

Obviously, the effective theory is needed for this derivation.  For
two-color QCD it was constructed previously \cite{Kanazawa:2009ks},
while for the isospin and adjoint cases it was constructed in
\cite{Kanazawa:PhD,Kanazawa:2012zr}.  Several points are worth noting
in this respect: (a) At high density the explicit breaking of
$\U(1)_A$ is strongly suppressed because instantons are screened.
Instead, $\U(1)_A$ is broken spontaneously by the diquark
condensate. (b) For chemical potential $\mu\gg\Lambda_\text{QCD}$ the
diquark pairing is of the BCS type since there is an attractive
channel between quarks near the Fermi surface.  (c) Because the
fermionic measure is positive definite, QCD inequalities allow us to
exclude certain symmetry-breaking patterns.  (d) The coefficient of
the mass term in the effective theory is not a free parameter but can
be computed in high-density effective theory (HDET)
\cite{Hong:1998tn}.

\subsection{Two-color QCD}

For simplicity we start with $N_f=2$ flavors.  The partition function
of two-color QCD is then
\begin{align}
  Z^{(N_f=2)}&=\big\langle\det(D+m_1)\det(D+m_2)\big\rangle_\text{YM}
  =\big\langle\det(D+m_1)\det(D^\dagger+m_2)\big\rangle_\text{YM}\,,
\end{align}
where $D=D(\mu)=\gamma_\nu D_\nu+\mu\gamma_4$ is the Euclidean Dirac
operator, the average is with respect to the Yang-Mills action, and in
the last step we have used the anti-unitary symmetry
$[C\gamma_5\tau_2K,D]=0$ that is specific to gauge group $\SU(2)$.
(In this relation, $C$ is the charge conjugation operator, $\tau_2$ is
the second generator of the gauge group, and $K$ is the complex
conjugation operator.) To have a positive-definite measure we could
either choose $m_1=m_2=m$, which leads to some mathematical subtleties
\cite{Kanazawa:2012zr}, or $m_1=z$ and $m_2=z^*$ with $z\in\mathbbm
C$.  We make the latter choice.  In terms of the Dirac eigenvalues we
then have
\begin{align}
  Z^{(N_f=2)}=\Big\langle\prod_n\big[(\lambda_n+z)
  (\lambda_n^*+z^*)+\eps^2\big]\Big\rangle_\text{YM}\,,
\end{align}
where we have introduced a regulator $\eps$ for later use.  Taking
derivatives first and the thermodynamic and chiral limits (in this
order) thereafter yields
\begin{align}
  \label{eq:Z2}
  \lim_{z\to 0}\lim_{V_4\to \infty}
  \frac{1}{V_4}\,\frac{\partial^2 \log Z}{\partial z\partial z^*}
  =\int d^2\lambda\,\frac{\eps^2}{(|\lambda|^2+\eps^2)^2}\rho(\lambda)
  +\int d^2\lambda \int d^2\lambda'\,\frac{\lambda^*}{|\lambda|^2+\eps^2}
  \frac{\lambda'}{|\lambda'|^2+\eps^2}\,{\rho}_2^c(\lambda,\lambda')
\end{align}
with spectral density and connected two-point function defined by
\begin{align}
  \label{eq:dens}
  \rho(\lambda) &= \lim_{z\to 0}\lim_{V_4\to\infty}\frac1{V_4}
  \Big\langle{\sum_n
    \delta^2(\lambda-\lambda_n)}\Big\rangle_{N_f=2}\,,\\
  \rho_2^c(\lambda,\lambda')&=\lim_{z\to 0}\lim_{V_4\to\infty}
  \bigg[\frac1{V_4} 
  \Big\langle{\sum_m \delta^2(\lambda\!-\!\lambda_m)
    \sum_n \delta^2(\lambda'\!-\!\lambda_n)}\Big\rangle_{\!N_f=2}
  \!-\! V_4\rho(\lambda)\rho(\lambda')\bigg].
\end{align}
The last term in \eqref{eq:Z2} vanishes due to chiral symmetry,
${\rho}^c_2(\lambda,\lambda')={\rho}^c_2(-\lambda,\lambda')
={\rho}^c_2(\lambda,-\lambda')={\rho}^c_2(-\lambda,-\lambda')$, and
using the delta-function $\delta^2(z) = \lim_{\eps\to 0}
\eps^2/\pi(|z|^2+\eps^2)^2$ in the complex plane we obtain
\begin{align}
  \label{eq:Z2diff}
  \lim_{\eps\to 0} \lim_{z\to 0} \lim_{V_4\to \infty}
  \frac{1}{V_4}\,\partial_{z^*}\partial_z \log Z^{(N_f=2)}
  = \pi{\rho}(0)\,.
\end{align}

From the effective theory of \cite{Kanazawa:2009ks}, with small
technical modifications to allow for complex $M=\diag(z,z^*)$, we
obtain for the shift in the free energy due to the BCS pairing of
quarks near the Fermi surface
\begin{align}
  \label{eq:dE2}
  \delta\mathcal E=\min_{A\in\U(1)}\left\{-\frac3{2\pi^2}\Delta^2
    (A^2+A^{*2})\det M\right\}\,.
\end{align}
For $M=\diag(z,z^*)$ the minimum is obtained for $A=\pm1$, which
yields
\begin{align}
  \frac1{V_4}\,\log Z^{(N_f=2)}=\frac3{\pi^2}\Delta^2zz^*
  +H_2(z^2+z^{*2})+O(|z|^3)\,,
\end{align}
where the second term on the RHS comes from a high-energy term $H_2\tr
M^2$.  (At high density the coefficient $H_2$ can be computed
perturbatively \cite{Kanazawa:2012zr}.) Differentiating gives
\begin{align}
  \label{eq:Z2effdiff}
  \lim_{z\to 0} \lim_{V_4\to \infty}\frac{1}{V_4}
  \,\partial_{z^*}\partial_z \log Z^{(N_f=2)}
  =\frac3{\pi^2}\Delta^2\,,
\end{align}
which combined with \eqref{eq:Z2diff} and $d_\text{rep}=2$ results in
\eqref{eq:BC}.  There are some technical subtleties in this derivation
for which we refer to \cite{Kanazawa:2012zr}.

Let us now generalize this to an arbitrary even number of flavors.
(We cannot say anything about odd $N_f$ since the measure is not
positive definite then.)  We choose the mass matrix to be
$M=\diag(z,\ldots,z,z^*,\ldots,z^*)$ with $N_f/2$ entries each of $z$
and $z^*$.  For the fundamental theory everything goes through as
before, and instead of \eqref{eq:Z2diff} we now obtain
\begin{align}
  \label{eq:Zdiff}
  \lim_{\eps\to 0} \lim_{z\to 0} \lim_{V_4\to \infty}
  \frac{1}{V_4}\,\partial_{z^*}\partial_z \log Z
  = \frac{N_f}2\pi{\rho}(0)\,.  
\end{align}
In the effective theory we obtain from the
results of \cite{Kanazawa:2009ks}
\begin{align}
  \delta\mathcal E=-\frac{3\Delta^2}{4\pi^2}\max_{A,\Sigma_L,\Sigma_R}
  \re \Big\{ A^2\tr(M\Sigma_RM^T\Sigma_L^\dagger) +
  A^{*2}\tr(M\Sigma_LM^T\Sigma_R^\dagger) \Big\}
\end{align}
with $A\in\U(1)$ and $\Sigma_{L,R}\in\SU(N_f)/\Sp(N_f)$.  Note that
additional terms are allowed by symmetries, but their coefficients are
zero in HDET.  The maximum is obtained for $A=\pm1$ and
\begin{align*}
  \Sigma_{L,R}=\pm
  \begin{pmatrix}0&-\1_{N_f/2}\\\1_{N_f/2}&0\end{pmatrix},
\end{align*}
which gives 
\begin{align}
  \delta\mathcal E=-\frac{3N_f}{2\pi^2}\Delta^2zz^*\,.
\end{align}
Taking derivatives w.r.t.\ $z$ and $z^*$ results in
\eqref{eq:Z2effdiff} with an extra factor of $N_f/2$ on the RHS.
Hence we again obtain \eqref{eq:BC}.

\subsection{QCD at nonzero isospin density}

We now consider QCD with $N_c\ge2$ colors and $N_f=2$ flavors at
nonzero isospin chemical potential $\mu_I=2\mu$.  The partition
function of the fundamental theory is
\begin{align}
  Z=\big\langle\det(D(\mu)+m_1)\det(D(-\mu)+m_2)\big\rangle_\text{YM}
  =\big\langle\det(D+m_1)\det(D^\dagger+m_2)\big\rangle_\text{YM}
\end{align}
since $D(-\mu)=-D(\mu)^\dagger$ and the Dirac eigenvalues occur in
pairs $\pm\lambda$.  Choosing $M=\diag(z,z^*)$ again results in
\eqref{eq:Z2diff}.

To construct the effective theory we note that at large $\mu_I$ BCS
pairing of type $\ev{\bar d\gamma_5u}$ occurs near the Fermi surface
\cite{Son:2000xc,Son:2000by}.  The analysis is similar to two-color
QCD, but the coset space is now $\U(1)_A\times\U(1)_{I_3}$
\cite{Kanazawa:PhD}, where the latter is the $\U(1)$ symmetry with
respect to the third isospin generator.  For the free-energy shift we
obtain
\begin{align}
  \delta\mathcal E=-c_\text{iso}\Delta^2\max_{A\in\U(1)}
  \left\{(A^2+A^{*2})\det M\right\}
  =-2c_\text{iso}\Delta^2zz^*
\end{align}
with $A=\pm1$ in the last step.  $c_\text{iso}$ is a positive
coefficient that can be computed in HDET following
\cite{Schafer:2001za,Kanazawa:2009ks}.  Using a result of
\cite{Hanada:2011ju} we obtain $c_\text{iso}=3N_c/4\pi^2$, which gives 
\begin{align}
  \label{eq:iso}
  \frac{1}{V_4}\log Z(z,z^*)
  =\frac{3N_c}{2\pi^2} \Delta^2 zz^* + H_2 (z^2+z^{*2})+O(|z|^3)
\end{align}
and thus \eqref{eq:BC} with $d_\text{rep}=N_c$.

\subsection{Adjoint QCD}

We now to turn to QCD with fermions transforming in the adjoint
representation of $\SU(N_c)$.  We again consider an even number of
flavors\footnote{$N_f$ must be even if we want to use a mass term of
  the form $M=\diag(z,\ldots,z,z^*,\ldots,z^*)$.  With real masses we
  could also consider odd $N_f$ since for adjoint fermions the measure
  is positive definite in this case.  However, then the $\Delta^2$ and
  $H_2$ terms in \eqref{eq:adj} have the same mass dependence, i.e.,
  they are proportional to $m_1^2+m_2^2+\ldots$ This is because for
  real masses the maximum in \eqref{eq:dEadj} is obtained for
  $\Sigma_{L,R}=\pm\1_{N_f}$.  For two-color QCD this argument is not
  applicable since $\pm\1_{N_f}$ is not in the coset space
  $\SU(N_f)/\Sp(N_f)$.  E.g., for $N_f=2$ with real masses we have
  $\delta\mathcal E\propto m_1m_2$, see \eqref{eq:dE2}.} and
$M=\diag(z,\ldots,z,z^*,\ldots,z^*)$.  The partition function of the
fundamental theory is
\begin{align}
  Z=\big\langle{\det}^{N_f/2}(D+z){\det}^{N_f/2}(D+z^*)\big\rangle_\text{YM}
  =\big\langle{\det}^{N_f/2}(D+z){\det}^{N_f/2}(D^\dagger+z^*)
  \big\rangle_\text{YM}\,,
\end{align}
where in the last step we have used the anti-unitary symmetry
$[C\gamma_5K,D]=0$ of the Dirac operator with adjoint fermions.  In
the fundamental theory we thus obtain \eqref{eq:Zdiff} again.

The construction of the low-energy effective theory for this case
proceeds in analogy to \cite{Kanazawa:2009ks}, except that the coset
space is now $\U(1)_B\times\U(1)_A\times
[\SU(N_f)_L/\SO(N_f)_L]\times[\SU(N_f)_R/\SO(N_f)_R]$.  For the
free-energy shift we obtain with $A\in\U(1)$ and
$\Sigma_{L,R}\in\SU(N_f)/\SO(N_f)$
\begin{align}
  \label{eq:dEadj}
  \delta\mathcal E & =-c_\text{adj} \Delta^2\max_{A,\Sigma_L,\Sigma_R}
  \re \Big\{ A^2\tr( M\Sigma_RM^T\Sigma_L^\dag)
  + A^{*2}\tr(M\Sigma_LM^T\Sigma_R^\dagger) \Big\}\notag\\
  & = -c_\text{adj} \Delta^2\cdot2N_fzz^*\qquad\text{for}\quad
  A=\pm1\,,\quad \Sigma_{L,R}=\pm
  \begin{pmatrix}0&\1_{N_f/2}\\\1_{N_f/2}&0\end{pmatrix}\,.
\end{align}
Again, additional terms are allowed by symmetries, but their
coefficients are zero in HDET.  To compute the coefficient
$c_\text{adj}$ in HDET we note that diquark condensation again occurs
in the pseudoscalar channel \cite{Kanazawa:PhD} but that the diquark
condensate is now symmetric in the color and flavor indices (and
antisymmetric in the spin indices as before).  Choosing the phases of
the right- and left-handed condensates so as to minimize the
ground-state energy we obtain after some calculations
$c_\text{adj}=3(N_c^2-1)/8\pi^2$ and thus\footnote{Note that the
  factor of $N_f/2$ in the $H_2$ term was erroneously omitted in
  \cite[eq.~(60)]{Kanazawa:2012zr}.}
\begin{align}
  \label{eq:adj}
  \frac{1}{V_4}\log Z(z,z^*) = \frac{3(N_c^2-1)}{4\pi^2} N_f \Delta^2
  zz^* + H_2\frac{N_f}2 (z^2+z^{*2})+O(|z|^3)\,.
\end{align}
Taking derivatives w.r.t.\ $z$ and $z^*$ again results in
\eqref{eq:BC}, now with $d_\text{rep}=N_c^2-1$.

\section{Consistency with microscopic limit and random matrix theory}

Let us define the spectral density in a finite volume in analogy to
\eqref{eq:dens},
\begin{align}
  \bar\rho(\lambda) = \lim_{z\to 0}\frac1{V_4}
  \Big\langle{\sum_n\delta^2(\lambda-\lambda_n)}\Big\rangle_{N_f}\,,
\end{align}
where $z\to0$ indicates that we only consider the chiral limit.  To
resolve individual eigenvalues near zero one defines the so-called
microscopic spectral density $\rho_s$ by rescaling the eigenvalues
with an appropriate power of the four-volume, which in our case is
$\sqrt{V_4}$.  It is well known in many other cases that $\rho_s$ is a
universal function that only depends on the symmetries and the
symmetry-breaking pattern of the system.  Therefore it can very
economically be computed in the simplest theory respecting these
symmetries, which is random matrix theory (RMT).  RMT can also be
understood as a zero-dimensional limit of the effective theory.  To
match results from the physical theory with RMT results a physical
scale has to be set.  This can be done, e.g., by matching the
mass-dependence of the partition functions of the effective theory and
of RMT.  In the following we show, for all three cases we considered,
that the correct rescaling is given by
\begin{align}
  \label{eq:rhos}
  \rho_s(\xi) = \lim_{V_4\to\infty}
  \frac{2\pi^2}{3d_\text{rep}\Delta^2}\,\bar\rho \left(
    \sqrt{\frac{2\pi^2}{3d_\text{rep}V_4\Delta^2}}\ \xi \right)
\end{align}
and that the asymptotic behavior of the RMT result for $\rho_s(\xi)$
is consistent with \eqref{eq:BC}.

Let us begin with two-color QCD.  In this case \eqref{eq:rhos} with
$d_\text{rep}=2$ was already proposed in \cite{Kanazawa:2009ks}.  The
corresponding RMT was constructed in \cite{Kanazawa:2009en}, and the
analytical RMT result for $\rho_s(\xi)$ was computed in
\cite{Akemann:2010tv}.  From this result one finds
\begin{align}
  \label{eq:limit}
  \lim_{|\xi|\to\infty}\rho_s(\xi)=\frac1\pi\,.
\end{align}
Note that \eqref{eq:rhos} is also valid for $|\xi|\to\infty$ but that
on the RHS the thermodynamic limit is taken first.  Hence
\eqref{eq:rhos} and \eqref{eq:limit} imply
\begin{align}
  \label{eq:cons}
  \frac{2\pi^2}{3d_\text{rep}\Delta^2}\,\bar\rho(0)=\frac1\pi\,,
\end{align}
which agrees with \eqref{eq:BC}.  Note that this should not be
considered as a derivation of \eqref{eq:BC} but as a consistency
check.

For QCD with isospin chemical potential the mapping of the physical
theory to RMT was derived in \cite[Sec.~4.2]{Kanazawa:PhD}, except
that a prefactor still needed to be computed that can be fixed by
matching \eqref{eq:iso} and \cite[Eq.~(4.37)]{Kanazawa:PhD}.  Using
this prefactor in \cite[Eq.~(4.49)]{Kanazawa:PhD} then gives
\eqref{eq:rhos} with $d_\text{rep}=N_c$.  The analytical RMT result
for $\rho_s(\xi)$ has been computed previously
\cite[Eq.~(4.51)]{Kanazawa:PhD}.  Adapting this result to our case
shows that \eqref{eq:limit} holds so that we again obtain the
consistency check \eqref{eq:cons}.

For adjoint QCD the mapping of the physical theory to RMT was derived
in \cite[Sec.~4.3]{Kanazawa:PhD}, again with the exception of a
numerical prefactor.  The microscopic spectral density for this case
can be obtained as a certain limit of a more general result
\cite{Akemann:2005fd}.  The missing prefactor is obtained by matching
\eqref{eq:adj} for $N_f=2$ with \cite[Eq.~(4.17)]{Akemann:2005fd}.
Using this prefactor in \cite[Eq.~(4.9)]{Akemann:2005fd} again gives
\eqref{eq:rhos}, now with $d_\text{rep}=N_c^2-1$.  From the analytical
RMT result for $\rho_s(\xi)$ (modulo a factor of 4 due to different
conventions) we can again conclude that \eqref{eq:limit} and thus also
\eqref{eq:cons} holds.

\section{Conclusions}

We have shown that three QCD-like theories at high density obey a
Banks-Casher-type relation that relates the BCS gap to the
two-dimensional Dirac density at the origin.  The relation is formally
identical for all three cases and only differs in the dimension of the
gauge group representation in which the fermions transform. To derive
the Banks-Casher-type relation we also constructed the hitherto
unknown low-energy effective theories for two of the cases.  In the
derivation of the Banks-Casher-type relation it is essential that the
fermionic measure is positive definite.  Therefore we cannot
generalize our result to the color-superconducting state of QCD at
high density.  Presumably the connection between the Dirac spectrum
and the BCS gap is more complicated in this case, similar to the
situation at low density \cite{Osborn:2005ss,Osborn:2008jp}.

Our results, together with other known results for microscopic
spectral correlations and spectral sum rules, can in principle be
checked in lattice simulations since there is no sign problem.
However, it is difficult to reach densities high enough for our
analysis to be valid and still keep the lattice spacing reasonably
small.

\bibliographystyle{JBJHEP}
\bibliography{new_bc}

\end{document}